\documentclass[twocolumn,showpacs,preprintnumbers,superscriptaddress,amsmath,amssymb,prl]{revtex4}

\usepackage{graphicx}
\usepackage{dcolumn}
\usepackage{bm}
\usepackage{color}

\begin{document}

\preprint{}

\title{Nature of Well-Screened State in Hard X-ray Mn 2$p$ Core-Level Photoemission of La$_{1-x}$Sr$_x$MnO$_3$ Films}

\author{K.~Horiba}
\email{horiba@spring8.or.jp}
\affiliation {Soft X-Ray Spectroscopy Laboratory, RIKEN/SPring-8, Mikazuki-cho, Hyogo 679-5148, Japan}

\author{M.~Taguchi}
\affiliation {Soft X-Ray Spectroscopy Laboratory, RIKEN/SPring-8, Mikazuki-cho, Hyogo 679-5148, Japan}

\author{A.~Chainani}
\affiliation {Soft X-Ray Spectroscopy Laboratory, RIKEN/SPring-8, Mikazuki-cho, Hyogo 679-5148, Japan}

\author{Y.~Takata}
\affiliation {Soft X-Ray Spectroscopy Laboratory, RIKEN/SPring-8, Mikazuki-cho, Hyogo 679-5148, Japan}

\author{E.~Ikenaga}
\affiliation {JASRI/SPring-8, Mikazuki-cho, Hyogo 679-5198, Japan}

\author{D.~Miwa}
\affiliation {Coherent X-Ray Optics Laboratory, RIKEN/SPring-8, Mikazuki-cho, Hyogo 679-5148, Japan}

\author{Y.~Nishino}
\affiliation {Coherent X-Ray Optics Laboratory, RIKEN/SPring-8, Mikazuki-cho, Hyogo 679-5148, Japan}

\author{K.~Tamasaku}
\affiliation {Coherent X-Ray Optics Laboratory, RIKEN/SPring-8, Mikazuki-cho, Hyogo 679-5148, Japan}

\author{M.~Awaji}
\affiliation {JASRI/SPring-8, Mikazuki-cho, Hyogo 679-5198, Japan}

\author{A.~Takeuchi}
\affiliation {JASRI/SPring-8, Mikazuki-cho, Hyogo 679-5198, Japan}

\author{M.~Yabashi}
\affiliation {JASRI/SPring-8, Mikazuki-cho, Hyogo 679-5198, Japan}

\author{H.~Namatame}
\affiliation {HiSOR, Hiroshima University, Higashi-Hiroshima, Hiroshima 739-8526, Japan}

\author{M.~Taniguchi}
\affiliation {HiSOR, Hiroshima University, Higashi-Hiroshima, Hiroshima 739-8526, Japan}

\author{H.~Kumigashira}
\affiliation {Department of Applied Chemistry, The University of Tokyo, Bunkyo-ku, Tokyo 113-8656, Japan}

\author{M.~Oshima}
\affiliation {Department of Applied Chemistry, The University of Tokyo, Bunkyo-ku, Tokyo 113-8656, Japan}

\author{M.~Lippmaa}
\affiliation {Institute for Solid State Physics, The University of Tokyo, Kashiwa, Chiba 277-8581, Japan}

\author{M.~Kawasaki}
\affiliation {Institute for Materials Research, Tohoku University, Sendai, Miyagi 980-8577, Japan}

\author{H.~Koinuma}
\affiliation {Materials and Structures Laboratory, Tokyo Institute of Technology, Yokohama, Kanagawa 226-8503, Japan}

\author{K.~Kobayashi}
\affiliation {JASRI/SPring-8, Mikazuki-cho, Hyogo 679-5198, Japan}

\author{T.~Ishikawa}
\affiliation {Coherent X-Ray Optics Laboratory, RIKEN/SPring-8, Mikazuki-cho, Hyogo 679-5148, Japan}

\author{S.~Shin}
\affiliation {Soft X-Ray Spectroscopy Laboratory, RIKEN/SPring-8, Mikazuki-cho, Hyogo 679-5148, Japan}
\affiliation {Institute for Solid State Physics, The University of Tokyo, Kashiwa, Chiba 277-8581, Japan}

\date{\today}

\begin{abstract}
Using hard x-ray (HX; $h\nu$~=~5.95~keV) synchrotron photoemission spectroscopy (PES), we study the intrinsic electronic structure of La$_{1-x}$Sr$_x$MnO$_3$ (LSMO) thin films. Comparison of Mn~2$p$ core-levels with Soft x-ray (SX; $h\nu$~$\sim$~1000~eV) -PES shows a clear additional well-screened feature only in HX-PES. Take-off-angle dependent data indicate its bulk ($\ge$~20~{\AA}) character. The doping and temperature dependence track the ferromagnetism and metallicity of the LSMO series. Cluster model calculations including charge transfer from doping induced states show good agreement, confirming this picture of bulk properties reflected in Mn~2$p$ core-levels using HX-PES. 
\end{abstract}

\maketitle

Hole-doped manganese oxides with a perovskite structure of $Re_{1-x}Ae_x$MnO$_3$ ($Re$ and $Ae$ being trivalent rare earth : Nd, Pr, Sm, etc. and divalent alkaline earth elements : Ca, Sr, Ba, respectively) exhibit a rich phase diagram originating in complex collective phenomena due to interplay among spin, charge, orbital, and lattice degrees of freedom \cite{MIT_Rev, CMR_Book}. Among the manganites, La$_{1-x}$Sr$_x$MnO$_3$ (LSMO) is a prototypical series showing the largest one-electron bandwidth and accordingly, is less significantly affected by electron-lattice and Coulomb correlation effects \cite{CMR_Book}. The parent compound LaMnO$_3$ is an antiferromagnetic (AFM) insulator which becomes, on hole-doping induced by substitution of Sr for La, a ferromagnetic (FM) metal \cite{Urushibara} exhibiting colossal magnetoresistance (CMR). The optimal doped compound ($x$~=~0.4) exhibits the highest Curie temperature ($T_C$) of 360~K among manganites and a half-metallic nature \cite{JHPark}. Further hole-doping induces a magnetic transition, transforming the FM metal to an AFM metal phase for $x$~{\textgreater}~0.5  \cite{Fujishiro}. In the case of thin films, the critical temperature and resistivity change slightly compared to the bulk materials due to the strain from the substrate, but the qualitative physical properties are similar to the bulk materials, provided the films are at least $\sim$~10 unit cells ($\sim$~30~{\AA}) thick \cite{Izumi_APL, Izumi_PRB, Fukumura, Horiba_D}.

In particular, high-quality bulk and thin films of the LSMO series do not exhibit charge order and are also free of micro- and nano-scale phase separation phenomena seen in the La-Ca, Nd-Sr and Pr-Sr manganites \cite{CMR_Book}. However, ultra thin films of LSMO (i.e. {\textless}~30~{\AA} or 10 unit cell thickness) are known to show a suppression of metallicity, ferromagnetic $T_C$ and magnetization \cite{Izumi_APL, Izumi_PRB}. In order to clarify the origin of these unusual physical properties, it is important to investigate the electronic structure of LSMO with a depth sensitive probe. Photoemission spectroscopy (PES) has long played a central role in studying the electronic structure of strongly correlated electron systems including manganese oxides \cite{Horiba_D, Chainani, Saitoh1, Sarma, Matsuno, Chuang}. Temperature dependent half-metallic ferromagnetism, charge and orbital ordering, and its connection with the electronic structure and colossal magnetoresistance of the manganites have been clarified \cite{Saitoh1, Sarma, Chuang}. Nevertheless, the change in the Mn~2$p$ spectra of manganese oxides with hole doping is still not conclusive \cite{Horiba_D, Chainani, Saitoh1, Matsuno}. Core-level spectra,  such as the 2$p$ levels of manganese, are very surface sensitive even if soft x-rays (SX; $h\nu$~$\sim$~1000~eV) are used, owing to short mean free paths of the emitted low kinetic energy electrons. In addition,  since correlation-induced changes at the surface of 3$d$ valence electron systems has been reported \cite{Maiti, Mo, Sekiyama}, the 2$p$ core-levels of 3$d$ transition metal compounds can also be expected to be much affected by the surface. Recently, high-resolution PES studies using hard x-rays (HX;  $h\nu$ $\sim$ 6 keV) have been realized \cite{Kobayashi, Takata}, making it possible to reveal the true bulk electronic structure upto depths of 50 to 100~{\AA}. In this work, we have performed Mn~2$p$ core-level spectroscopy of LSMO thin films using HX-PES. Composition, temperature, and depth-dependent experiments and cluster model calculations are applied to show that bulk properties are reflected in core-level HX-PES.

\begin{figure}
\includegraphics[width=0.8\linewidth]{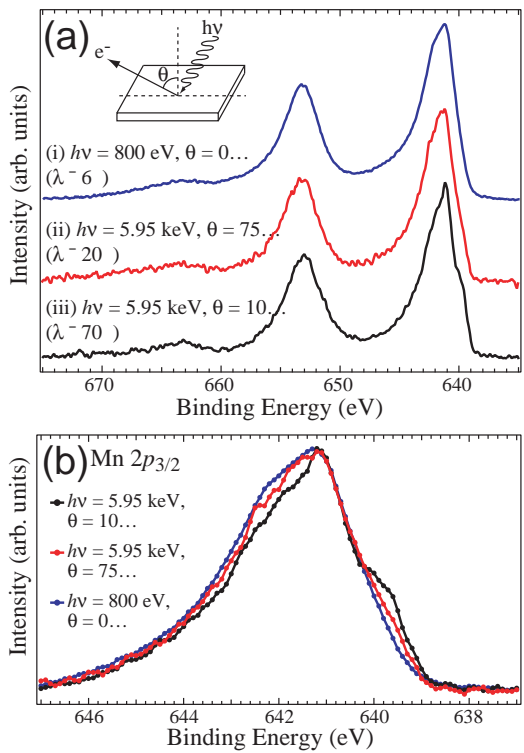}
\caption{\label{figure1} (Color) (a) Mn~2$p$ core-level spectra of LaMnO$_3$ measured with different probing depth by changing photon energy (800~eV and 5.95~keV) and emission angle (0$^{\circ}$, 10$^{\circ}$ and 75$^{\circ}$). The inset shows the experimental configuration. (b)~The Mn~2$p_{3/2}$ region on an expanded scale.}
\end{figure}

The LSMO thin films were grown epitaxially on SrTiO$_3$ (STO) substrates by laser molecular beam epitaxy (laser MBE) and the thickness was accurately determined from reflection high energy electron diffraction intensity oscillations to be 100 monolayers ($\approx$~400~{\AA}). Resistivity and magnetic properties of fabricated films were measured before PES measurements and the suitability of these values was confirmed \cite{Izumi_APL, Izumi_PRB, Fukumura, Horiba_D}. HX-PES experiments were carried out at undulator beamlines BL29XUL \cite{Tamasaku} and BL47XU of SPring-8. The detailed growth conditions and HX-PES instrumentation details are described in refs.~\onlinecite{Horiba_D, Horiba_RSI} and refs.~\onlinecite{Kobayashi, Takata}, respectively. The films were transferred into the PES chamber for HX-PES from air without any surface cleaning procedures. Hard x-rays as an excitation source enable us to measure PES spectra without surface cleaning procedures, due to the large escape depth of photoelectrons with high kinetic energy \cite{Kobayashi, Takata, Tanuma}. The total energy resolution was set to about 300~meV. The energy scale was calibrated using the the peak position of Au~4$f$ core-level and the Fermi level ($E_F$). SX-PES measurements were carried out at BL-2C of KEK-PF using a combined laser MBE and PES spectrometer \cite{Horiba_D, Horiba_RSI}.

\begin{figure}
\includegraphics[width=0.8\linewidth]{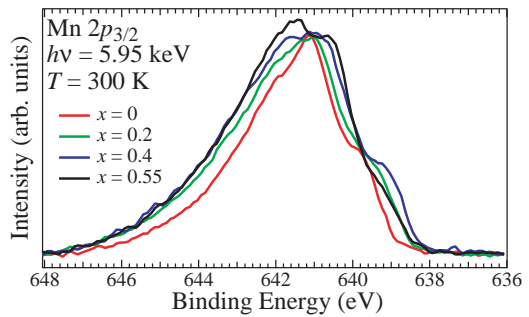}
\caption{\label{figure2} (Color) Hole concentration dependence of Mn~2$p_{3/2}$ spectra measured at 300~K.}
\end{figure}

Figure~\ref{figure1} shows Mn~2$p$ core-level spectra of LaMnO$_3$ at 300~K measured with different probing depths by changing the photon energy and emission angle. An integral background has been subtracted from raw data. The estimated probing depths \cite{Tanuma} at (i)~$h\nu$~=~800~eV, $\theta$~=~0$^{\circ}$ (SX-spectra), (ii)~$h\nu$~=~5.95~keV, $\theta$~=~75$^{\circ}$, and (iii)~$h\nu$~=~5.95~keV, $\theta$~=~10$^{\circ}$ (HX-spectra) are 6~{\AA}, 20~{\AA}, and 70~{\AA}, respectively. Note that the SX-PES spectrum has been measured using an {\it in~situ} technique \cite{Horiba_RSI} and surface contamination is negligible in the SX-PES spectrum. Nevertheless, there are obvious differences between the SX- and HX-PES spectra. In the SX-PES spectrum, a shoulder structure at the binding energy (BE) of about 642~eV is clearly observed. The intensity of this shoulder structure at the high BE side of the Mn~2$p_{3/2}$ systematically decreases with increasing probing depth. The systematic reduction of the shoulder structure indicates that this feature has a surface electronic structure component, which is minimized in the HX-PES spectrum.

A more remarkable difference between HX- and SX-PES spectra is a new shoulder structure at the low BE side of Mn~2$p_{3/2}$ main peak in the HX-PES spectra. The intensity of this feature systematically increases with increasing probing depth, in contrast to the high BE side shoulder of main peak. The data indicate that this feature is observed for depths $\ge$~20~{\AA} only. We have confirmed that Ti~2$p$ core-levels from the STO substrates cannot be observed using HX-PES with a probing depth of $\sim$~70~{\AA}, simply because the film thickness is $\approx$~400~{\AA}. This denies the possibility that the origin of this feature is the change in the electronic structure at the interface between the LSMO thin films and the STO substrates. Therefore, we conclude that this low BE feature is derived from the bulk electronic structure.  This bulk-derived feature has not been observed in SX-PES measurements \cite{Horiba_D, Chainani, Saitoh1, Matsuno} and is only observed using the present higher probing depth HX-PES measurements. In the following, we study its doping and temperature dependence using HX-PES.

\begin{figure}
\includegraphics[width=0.95\linewidth]{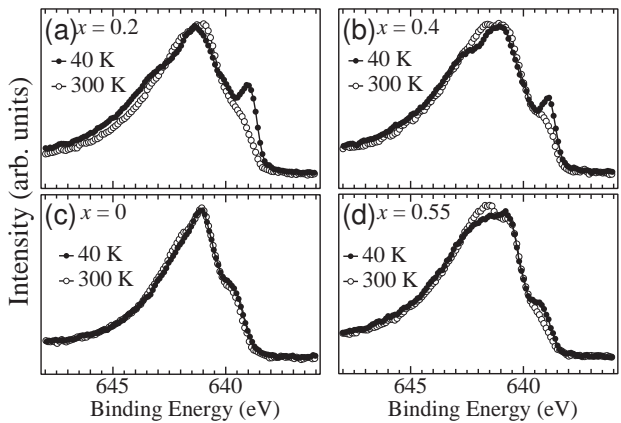}
\caption{\label{figure3} Temperature dependence of Mn~2$p_{3/2}$ spectra with various hole concentrations, (a)~$x$~=~0.2, (b)~$x$~=~0.4, (c)~$x$~=~0, (d)~$x$~=~0.55.}
\end{figure}

The Mn~2$p_{3/2}$ HX-PES spectra show dramatic changes with hole-doping, as shown in Fig.~\ref{figure2}. The high BE side of the main peak increases systematically with hole-doping. This structure is attributed to the Mn$^{4+}$ derived state, since it increases systematically with hole-doping and is best seen in the $x$~=~0.4 and 0.55 compositions. Separating out the Mn$^{4+}$ contribution from the spectra, the main peak position of Mn$^{3+}$ shifts to lower BE systematically with hole-doping, suggesting a rigid-band-like shift of the Mn~2$p$ core-levels and absence of phase separation tendencies in the LSMO thin film system \cite{Horiba_D, Matsuno}. The bulk-derived feature at low BE side of main peak also shows obvious changes with hole-doping and is discussed in detail below. None of the changes observed in HX-PES Mn~2$p_{3/2}$ spectra have been observed in the SX-PES spectra of bulk poly- and single-crystals as well as thin films of manganites \cite{Horiba_D, Chainani, Saitoh1, Matsuno}, probably owing to modifications of the electronic structure within about 20~{\AA} from the surface and the surface sensitivity of SX-PES.

The bulk-derived low BE or "well-screened" feature exhibits the following characteristics: the separation between the main peak and the well-screened feature increases with hole-doping until $x$~=~0.4, but reduces for $x$~=~0.55. This behavior is similar to the physical properties, that is, with increasing $x$, the hole-doping produces a FM phase with increasing T$_C$ and reduced resistivity until $x$~=~0.4 \cite{Urushibara}. On further hole-doping, a magnetic transition from the FM metal to AFM metal state is induced for $x$~{\textgreater}~0.5 \cite{Fujishiro}. In order to confirm this relation between the bulk electronic structure and the physical properties, we have checked the temperature dependence of the well-screened feature. Figure~\ref{figure3} shows the comparison between the Mn~2$p_{3/2}$ spectra measured at 300~K and 40~K. In the  $x$~=~0.2 and $x$~=~0.4 spectra, the well-screened feature exhibits drastic increase of intensity and becomes a sharp peak structure. On the other hand, the low BE feature of $x$~=~0 and $x$~=~0.55 shows little change in the intensity and the spectral shape.

\begin{figure}
\includegraphics[width=0.85\linewidth]{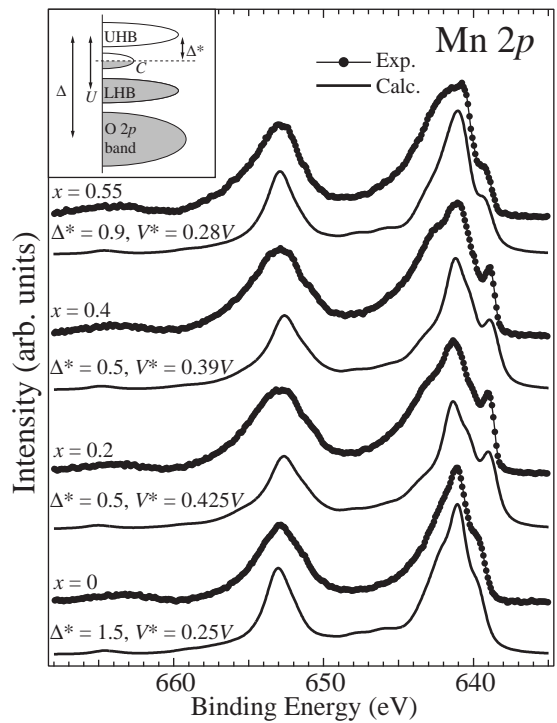}
\caption{\label{figure4} Comparison between the cluster calculation and HX-PES spectra of Mn~2$p$ core-level. The inset shows a schematic diagram of energy levels on the valence band.}
\end{figure}

Concerning the temperature-dependent physical properties of LSMO thin films, $x$~=~0.2 compound shows an insulator-to-metal transition between 300~K and 40~K \cite{Horiba_D}. The $x$~=~0.4 compound shows metallic behavior at all temperatures below 300~K, but the metallicity increases on stabilizing the FM state at low temperature, and is attributed to its half-metallic nature \cite{JHPark}. On the other hand, while the $x$~=~0 stoichiometric compound is insulating, it also is effectively hole doped due to excess oxygen. As is well-known, excess oxygen is easily introduced during the growth of thin films, and the electronic structure is significantly changed by the existence of excess oxygen, as has been reported using O~1$s$ x-ray absorption spectra of LaMnO$_{3+\delta}$ \cite{JHPark_LCMO}. For the as-grown or oxygen-annealed samples, a composition of LaMnO$_{3+\delta}$, with $\delta$ being about 0.1 (or 20~\% Mn$^{4+}$) has been reported \cite{Chainani,JHPark_LCMO}. For $x$~=~0.55, the material is in the AFM phase. Given the observations, we assume that the well-screened feature is strongly related to the doping-induced density of state (DOS) responsible for the ferromagnetism and metallicity. 

In order to confirm this assumption, we have tried to reproduce the spectra using MnO$_6$ (3$d^4$) cluster model calculations with $D_{4h}$ symmetry. In addition to the usual model \cite{Taguchi_SRL} which includes the Mn~3$d$ and ligand O~2$p$ states, we have introduced new states at $E_F$ as labeled $C$ in the inset of Fig.~\ref{figure4}. These new states represent the doping-induced states which develop into a metallic band at $E_F$, but are approximated as a level for simplicity, following earlier work \cite{imer87,Taguchi_condmat}. Very recently, a similar model using dynamic mean field theory has been successfully applied to calculate core-level spectra in a series of ruthenates across the metal-insulator transition, but in the absence of ligand states \cite{HDKim}. For the initial states, we use four configurations, namely 3$d^4$, 3$d^5\underline{\it L}$ where \underline{\it L} is a hole in the ligand O~2$p$ states, and 3$d^3C$ and 3$d^5\underline{\it C}$ which represent charge transfer (CT) between DOS at $E_F$ and Mn~3$d$ state. The cluster calculation is carried out for a high-spin configuration, consistent with the magnetic moment estimated from susceptibility measurements \cite{Urushibara}. We fit the experimental spectra by changing two parameters: the CT energy between Mn~3$d$ and the new $C$ states ($\Delta^*$) and the hybridization between Mn~3$d$ and the new $C$ states ($V^*$). Except for these two parameters, all other parameter values are fixed and determined from previous work \cite{Taguchi_SRL}: the $d$-$d$ Coulomb interaction of  Mn~3$d$ states $U$~=~5.1~eV, the CT energy between Mn~3$d$ and ligand O~2$p$ states $\Delta$~=~4.5~eV, the hybridization between Mn~3$d$ and ligand O~2$p$ states $V$~=~2.94~eV, the crystal field splitting 10$Dq$~=~1.5~eV, and the Coulomb interaction between Mn~3$d$ and Mn~2$p$ core hole states $U_{dc}$~=~5.4~eV. Figure~\ref{figure4} shows the comparison between the HX-spectra and the optimized calculations. For all $x$ values, the calculation reproduces well the intensity and position of the well-screened feature of HX-PES spectra. The well-screened feature in the calculation is anlaysed to originate from the 2$p^5$3$d^5\underline{\it C}$ configuration of the final state, and increases in intensity with increasing $V^*$. The cluster calculations indicate a larger hybridization strength $V^*$ with the coherent states, or increase in delocalization, for the FM compositions ($x$~=~0.2 and 0.4) as compared to the AFM compositions ($x$~=~0 and 0.55). This is consistent with the known half-metallic ferromagnetism which is stabilized with an increase in hybridization, for the manganites upon doping \cite{JHPark, FMCalc}. This also suggests an analogy with the Kondo coupling between $f$ states and conduction band states with $V^*(E_F)\propto\sqrt{D(E_F)}$ \cite{Handbook_RareEarth}, where $D(E_F)$ is DOS at $E_F$. It is also noted that the high BE side of Mn~2$p_{3/2}$ main peak, particularly for high doping, does not match with the calculations. This disagreement is due to the Mn$^{4+}$ derived state appearing at the high BE side of main peak with hole-doping, and is not included in the calculations.

In conclusion, we have performed bulk-sensitive PES on Mn~2$p$ core-level of LSMO thin films using Hard x-rays ($h\nu$ = 5.95 keV) and find a well-screened feature of the Mn~2$p_{3/2}$ peak. The feature shows a noticeable increase with decreasing temperature for the FM metal compositions, indicating that the origin of this feature is strongly related to the FM metal phase of LSMO thin films. A cluster model calculation including CT from doping induced DOS at $E_F$ to Mn~3$d$ states reproduce the doping dependence of the well-screened feature, indicating that the metallic DOS at $E_F$ is reflected in the Mn~2$p$ core-levels using bulk-sensitive HX-PES. Core level HX-PES can hence be reliably used to study the bulk character of correlated electron systems exhibiting insulator-metal and magnetic transitions and its relation with doping induced states, as well as other systems such as magnetic multilayers, buried interfaces, etc. up to depths of tens of nanometers.

\end{document}